\documentstyle[epsfig]{article}
\newcommand{\rb}{{r_{\ast}}}
\newcommand{\rt}{\tilde{r}}
\newcommand{\thetab}{{\theta_{\ast}}}

\newcommand{\BBox}{\Box}
\begin{document}

\begin{titlepage}

\begin{center}
{\large\bf Quasi-Spherical Light Cones of the Kerr Geometry}
\end{center}

\vfill

\begin{center}
Frans Pretorius and Werner Israel 
\end{center}
\vfill

\begin{center}{\sl
Canadian Institute for Advanced Research Cosmology Program \\
Department of Physics and Astronomy \\
University of Victoria \\
P.O. Box 3055 STN CSC \\
Victoria B.C., Canada V8W 3P6 \\
}\end{center}
\vfill

\begin{abstract} 
Quasi-spherical light cones are lightlike hypersurfaces of the Kerr 
geometry that are asymptotic to Minkowski light cones at infinity. We 
develop the equations of these surfaces and examine their properties. In 
particular, we show that they are free of caustics for all positive values of 
the Kerr radial coordinate $r$. Useful applications include the propagation of 
high-frequency waves, the definition of Kruskal-like coordinates for a 
spinning black hole and the characteristic initial-value problem. 
\end{abstract}

\begin{center}{\sl
PACS numbers: 0240, 0420
}\end{center}
\vfill

\end{titlepage}

\section{Introduction}

The Kerr geometry is a vacuum spacetime with a Weyl tensor of Petrov 
type D. According to the Goldberg-Sachs theorem \cite{wald}, it therefore 
possesses two congruences (ingoing and outgoing) of shear-free lightlike 
geodesics. Historically, these congruences played an essential role in the 
discovery of the Kerr solution \cite{kerr}, because the metric takes a 
simple explicit form in coordinates adapted to them.\par
In the original \cite{kerr},\cite{he} (Eddington-Kerr) form of the metric, 
the ingoing congruence consists of parametric curves of the Eddington-
Kerr coordinate $v$, often referred to as an ``advanced time''. This 
nomenclature can be misleading, because hypersurfaces of constant $v$ 
are actually spacelike, not lightlike, if the Kerr rotational parameter $a$ is 
nonzero. (The lightlike congruences are ``twisting", i.e., not orthogonal to 
any hypersurfaces, and hence not tangent to lightlike hypersurfaces if 
$a\neq 0$). \par
Thus, the Eddington-Kerr form of the metric, and the Boyer-Lindquist 
form \cite{he} derived from it by an elementary coordinate transformation, 
correspond to a ``threading'' \cite{bd} (i.e., a one-dimensional foliation) of 
the manifold by twisting lightlike geodesics. However, a ``slicing" (three 
dimensional foliation), in particular a light-like slicing, is for many 
purposes more advantageous and often corresponds more closely to the 
physics. \par
Light cones of the Kerr geometry have not previously been a subject of 
systematic study to our knowledge. Perhaps this stems in part from a 
feeling that such hypersurfaces would quickly develop caustics because of 
the twist inherent in the metric. Yet as characteristics these surfaces 
obviously play a key role in the physics of Kerr black holes, e.g in the 
propagation of wave fronts and high-frequency waves, and in 
characteristic initial value problems. \par
In this paper we first derive the general solution for axisymmetric lightlike 
hypersurfaces of the Kerr geometry (Sec. \ref{formal_sol}). We then focus 
on a particular foliation (invariant under time-displacement) by (either 
ingoing or outgoing) ``quasi-spherical" light cones, defined as degenerate 
3-spaces whose spatial sections are asymptotically spherical at infinity. 
These hypersurfaces become the standard Minkowski light cones when the 
Kerr mass parameter $m$ vanishes; if $a=0$, $m\neq0$, they are the 
familiar spherical light cones, $t \pm \rb = constant$, of Schwarzschild 
spacetime (where $\rb$ is the ``tortoise" coordinate). Quite generally, for 
arbitrary $m \geq 0$ and $a$, they have the remarkable property of being 
free of caustics for all positive values of the Kerr radial coordinate $r$. 
\par
These appealing features suggest that coordinates adapted to quasi-
spherical surfaces should be especially convenient and useful for the 
description of the Kerr geometry. The drawback (and it is considerable) is 
that the metric is no longer expressible in an explicit elementary form, 
because the new coordinates are elliptic functions of the Boyer-Lindquist 
coordinates. \par
We now briefly outline the contents of the paper. We begin in Sec. 2 by 
solving the eikonal equation for axisymmetric lightlike hypersufaces of the 
Kerr geometry. The general solution is expressed in terms of integrals of 
elliptic type, and involves an arbitrary function $f$ of one variable, to be 
fixed when the ``initial'' (e.g. asymptotic) shape of the surface is given. (If 
$m=0$ the geometry becomes flat and the integrals can be reduced to 
elementary form; in Sec. 3 we digress briefly to illustrate this.) Some 
general properties of axisymmetric lightlike surfaces are developed in Sec. 
4. In Sec. 5, guided by the results of Sec. 3 for Minkowski light cones, we 
make the special choice of $f$ appropriate for asymptotically spherical 
light cones in Kerr space. The key result, that these are free of caustics for 
$r > 0$, is demonstrated in Sec. 6. In Secs. 7 and 8 we show how the 
``quasi-spherical" coordinates $\rb$, $\thetab$ defined by these light cones 
can be used to construct high-frequency solutions to the wave equation and 
Kruskal-like coordinates for Kerr black holes. Sec. 9 summarizes the
limiting cases that can provide serviceable approximations for the elliptic functions 
which appear in the quasi-spherical form of the metric. Finally, in Sec. 10 
we present some results of numerical integrations for the inward evolution 
of the light cone generators. 

\section{Axisymmetric lightlike hypersurfaces}\label{formal_sol}

The equation of an arbitrary axially symmetric lightlike hypersurface of 
the Kerr geometry is expressible in terms of elliptic integrals as we now 
proceed to show.\par
We write the Kerr metric in its standard (Boyer-Lindquist) form

\begin{equation}\label{kerr}
 ds^2 = \frac{\Sigma}{\Delta} dr^2+ \Sigma d\theta^2 + R^2\sin^2\theta 
d\phi^2  -
\frac {4mar\sin^2\theta}{\Sigma} d\phi dt - \left(1-\frac 
{2mr}{\Sigma}\right)dt^2,
\end{equation}

where

\begin{equation}\label{util}
\Sigma = r^2+a^2\cos^2\theta,
\ \ R^2 = r^2+a^2+\frac{2ma^2r\sin^2\theta}{\Sigma},
\ \ \Delta = r^2+a^2-2mr,
\end{equation}

and we note the useful identities

\begin{equation}\label{ident}
\Sigma R^2=(r^2+a^2)^2-\Delta a^2 \sin^2\theta,
\end{equation}

\begin{equation}\label{det}
g_{\phi\phi}g_{tt}-g_{\phi t}^2=-\Delta \sin^2\theta.
\end{equation}

The equation 

\begin{equation}\label{cond}
 v (t,r,\theta)=t + \epsilon \rb(r,\theta)=constant, \ \ \epsilon = \pm 1
\end{equation}

represents an axisymmetric lightlike hypersurface (ingoing for 
$\epsilon=1$, outgoing for $\epsilon=-1$) if 
$g^{\alpha \beta}(\partial_\alpha v)(\partial_\beta v)=0$, i.e., if

\begin{equation}\label{pde}
 \Delta (\partial_r \rb)^2 + (\partial_{\theta}\rb)^2
=(r^2+a^2)^2/\Delta-a^2\sin^2\theta.
\end{equation}

It is easy to obtain a particular (separable) solution of (\ref{pde}) 
depending on two arbitrary constants (a ``complete integral'') by adding 
and subtracting an arbitrary separation constant $a^2\lambda$ on the right 
-hand side. Let us define 

\begin{equation}\label{pq}
P^2(\theta,\lambda)=a^2(\lambda-\sin^2\theta),  \ \ \
Q^2(r,\lambda,m)=(r^2+a^2)^2-a^2\lambda\Delta.
\end{equation}

(Note the useful identity

\begin{equation}\label{ident2}
Q^2+\Delta P^2 = \Sigma R^2,
\end{equation}

which follows from (\ref{ident}).) A complete integral

\begin{equation}\label{rho}
\rb = \rho (r, \theta, \lambda,m),
\end{equation}

of (\ref{pde}) is then obtained by integrating the exact differential

\begin{equation}\label{drb}
d\rb = (Q/\Delta)dr + P d\theta
\end{equation}

at fixed $\lambda$. When (\ref{drb}) is integrated, a second, additive 
integration constant appears which we shall denote as $a^2 f(\lambda)/2$, 
where $f$ is an arbitrary function. \par
We next proceed in the usual way to promote this complete integral, 
depending upon the arbitrary constants $\lambda$ and $f(\lambda)$, to a 
general solution involving an arbitrary function. \par

In (\ref{rho}), $\rho$ is a function of three independent variables $r$, 
$\theta$ and $\lambda$ (not counting $m$ and $a$), and a more complete 
expression for its differential is 

\begin{equation}\label{drho}
d\rho = (Q/\Delta)dr + P d\theta + (a^2/2) F d\lambda,
\end{equation}

where $F$ is the partial derivative 

\begin{equation}\label{F}
(a^2/2) F(r,\theta,\lambda,m)=\partial_\lambda \rho(r,\theta,\lambda,m).
\end{equation}

Its explicit form may be taken to be 

\begin{equation}\label{F2}
F(r,\theta,\lambda,m)= 
\int_0^{\theta} \frac{d\theta'}{P(\theta',\lambda)}
+\int_r^{\infty} \frac{dr'}{Q(r',\lambda)} +
f'(\lambda).
\end{equation}

Up to now, we have taken $\lambda$ to be a constant, i.e., $d\lambda=0$ 
in (\ref{drho}). But we achieve the same effect (i.e., (\ref{drho}) still 
reduces to (\ref{drb}) ) even when $d\lambda \neq 0$ provided we require 
that $F=0$. In other words: the function $\rb(r,\theta)$, given by 
(\ref{rho}) with $\lambda$ now a \emph{function} $\lambda(r,\theta)$, 
remains a solution of (\ref{pde}) provided its extra dependence on 
$r,\theta$ through $\lambda$ does not change the algebraic form of the 
differential (\ref{drb}). This will indeed be so if we impose the constraint 

\begin{equation}\label{constraint}
F(r,\theta,\lambda,m)=0.
\end{equation}

This condition fixes the dependence of $\lambda$ on $r,\theta$ for any 
given choice of $f(\lambda)$. Thus we now have a general solution, 
depending upon an arbitrary \emph{function} $f[\lambda(r,\theta)]$. \par
The explicit form of the general solution (\ref{rho}) is then   

\begin{eqnarray}\label{solution}
\rho(r,\theta,\lambda)=\int \frac{r^2+a^2}{\Delta(r,m)} dr + 
\int_r^{\infty} \frac{r'^2+a^2 - Q(r',\lambda,m)}{ \Delta(r',m) }dr' \\ +
\int_0^{\theta} d\theta' P(\theta',\lambda) 
+ \frac{1}{2}a^2f(\lambda), \nonumber
\end{eqnarray}

where the radial dependence has been arranged to ensure convergence of 
the definite integral at its upper limit. When performing the integrations in 
(\ref{F2}) and (\ref{solution}), we are allowed to treat $\lambda$ as 
merely a passive constant parameter -- it is in fact a function of the limits 
of integration $r$,$\theta$, not of the integration variables -- with the 
constraint (\ref{constraint}) imposed \emph{a posteriori}. It is possible, 
though not especially illuminating or useful, to express each of these 
integrals in terms of standard elliptic integrals. 

\section{The case $m=0$: light cones in Minkowski 
space}\label{case_m_0}

As a simple illustrative example (and because we shall need some of the 
results later), we specialize in this section to the case $m=0$. The Kerr 
line-element reduces to the metric of flat spacetime expressed in terms of 
oblate spheroidal coordinates $r,\theta$ -- related to Cartesians by

\begin{equation}\label{oblate}
x^2+y^2=(r^2+a^2)\sin^2\theta, \ \ \ z=r\cos\theta.
\end{equation}

The second integral in (\ref{F2}) can be reduced to the same general form 
as the first integral when $m=0$. Assume that we are in a domain where 
$\lambda(r,\theta)<1$. Setting 

\begin{equation}\label{Q0}
Q_0(r',\lambda)=Q(r',\lambda,m=0) = \sqrt{r'^2+a^2}\sqrt{r'^2+a^2(1-
\lambda)},
\end{equation}

and making the substution 

\begin{equation}\label{sub}
r'=a\sqrt{1-\lambda}\sin \chi'/ \sqrt{\lambda-\sin^2\chi'},
\end{equation}

we find

\begin{equation}
dr'/Q_0 = d\chi'/a\sqrt{\lambda-\sin^2\chi'}.
\end{equation}

Thus the two integrals in (\ref{F2}) can be combined into a single integral, 
to give

\begin{equation}\label{F_m0}
a F(r,\theta,\lambda,m=0)=\int_{\chi(r,\lambda)}^\theta 
\frac{d\chi'}{\sqrt{\lambda- \sin^2\chi'}} + g'(\lambda),
\end{equation}

where 

\begin{equation}\label{g}
g(\lambda)=a f(\lambda)+2\int_0^{\thetab} \sqrt{\lambda-\sin^2 \chi'} 
d\chi',
\end{equation}

and the form of $\chi(r,\lambda)$ is given by the ``unprimed'' version of 
(\ref{sub}); equivalently, 

\begin{equation}\label{theta_s}
\tan\thetab = \frac{\sqrt{r^2+a^2}}{r} \tan\chi, \ \ \  \lambda \equiv 
sin^2\thetab.
\end{equation}

The function $g(\lambda)$ is arbitrary. As an example, let us consider the 
simplest choice, $g(\lambda)=0$. The functional dependence 
$\lambda(r,\theta)$ corresponding to this choice is determined by the 
constraint $F=0$, which requires that $\chi(r,\lambda)=\theta$ by 
inspection of (\ref{F_m0}). Thus (\ref{theta_s}) gives

\begin{equation}\label{theta_s2}
\tan\thetab = \frac{\sqrt{r^2+a^2}}{r} \tan\theta = 
\frac{\sqrt{x^2+y^2}}{z}
\end{equation}

by (\ref{oblate}), showing that $\thetab$ is the spherical polar angle. With 
$\lambda(r,\theta)$ known from (\ref{theta_s2}), it is straightforward to 
integrate (\ref{drb}) to obtain

\begin{equation}\label{rs}
\rb=\sqrt{r^2+a^2\sin^2\theta}.
\end{equation} 

Thus $\rb$ is the usual spherical radius, and our solutions $v=t\pm 
\rb=constant$ of (\ref{pde}) are in this case the Minkowski light cones.

\section{Axisymmetric lightlike hypersurfaces: general properties} 
\label{gen_prop}

Returning to the general case, we shall now derive a number of results 
applicable to all axisymmetric lightlike hypersurfaces of Kerr. \par
Since the function $\lambda(r,\theta)$ is determined by the constraint 
$F=0$, the partial derivatives of $\lambda$ can be read off from $dF=0$ 
(\ref{F2}):

\begin{equation}\label{dl_mu}
\mu d\lambda=-dr/Q +d\theta/P, \ \ \ \mu \equiv -\partial F/\partial \lambda.
\end{equation}
\par
It follows from (\ref{drb}) and (\ref{dl_mu}) that $\nabla\rb \cdot 
\nabla\lambda =0$, i.e. that $\rb$ and $\lambda$ are orthogonal with 
respect to the intrinsic 2-metric 

\begin{equation}\label{2metric}
d\sigma^2 = \Sigma(dr^2/\Delta + d\theta^2)
\end{equation}

of the spatial sections $(\phi,t)=constant$ of the Kerr geometry. Since 
$\lambda$ is independent of $\phi$ and $t$, it follows further that 
\emph{$\lambda$ is constant along the lightlike generators}, i.e., 

\begin{equation}\label{lconst}
\ell^\alpha \partial_\alpha \lambda =0, \ \ \ \ell_\alpha = -\partial_\alpha v = 
-\partial_\alpha (t+\epsilon \rb).
\end{equation}   

If $\rb$ and $\lambda$ are adopted as coordinates in place of $r$ and 
$\theta$, the 2-metric (\ref{2metric}) becomes

\begin{equation}\label{2metric_rl}
d\sigma^2 = R^{-2}(\Delta d\rb^2 + L^2 d\lambda^2), \ \ \ L \equiv \mu P 
Q,
\end{equation}

where we have made use of (\ref{drb}), (\ref{dl_mu}) and (\ref{ident2}). 
The identity (\ref{det}) can be re-arranged in the form 

\begin{equation}\label{det2}
\left(1-\frac{2mr}{\Sigma}\right) + \omega_B^2 R^2 \sin^2\theta  = 
\frac{\Delta}{R^2},
\end{equation}

where

\begin{equation}\label{wb}
\omega_B=-\frac{g_{\phi t}}{g_{\phi \phi}} = \frac{2mar}{\Sigma R^2}
\end{equation}

is the Bardeen or ZAMO angular velocity $d\phi/dt$ which characterizes 
orbits having zero angular momentum. This enables us to recast the Kerr 
metric (\ref{kerr}) in the form

\begin{equation}\label{kerr_zamo}
ds^2=\frac{\Delta}{R^2}(d\rb^2-dt^2)+\frac{L^2}{R^2}d\lambda^2 + 
R^2\sin^2\theta (d\phi-\omega_B dt)^2.
\end{equation}

Thus 

\begin{equation}\label{grsrs}
g^{\rb\rb} = R^2/\Delta = -g^{tt},
\end{equation}

from which the null character of the 3-spaces $t \pm \rb = constant$ is 
directly evident. The degenerate intrinsic metric of these 3-spaces is 

\begin{equation}\label{dsll}
(ds^2)_{LL} = (L/R)^2 d\lambda^2 + R^2 \sin^2\theta(d\phi- \omega_B 
dt)^2,
\end{equation}

showing that the generators rotate with $ZAMO$ angular velocity relative 
to stationary observers at infinity. (This is directly obvious from 
(\ref{lconst}), which shows that $\ell_\phi=0$ for an axisymmetric 
hypersurface). From (\ref{dsll}) we see that caustics will develop when the 
degenerate volume element tends to zero, i.e. when $L \sin\theta 
\rightarrow 0$ (recalling that $\lambda$ has a fixed value along each 
generator).  \par
The integrability of (\ref{dl_mu}) for the exact differential $d\lambda$ 
provides a condition on the integrating factor $\mu$. This takes the form 
of an evolution equation for $\mu$ along the generators. A straightforward 
calculation, which calls upon (\ref{pq}), (\ref{ident2}), (\ref{F2}) and 
(\ref{dl_mu}) itself, yields

\begin{equation}\label{d_mu}
(\partial \mu /\partial r)_{\lambda}=\frac{a^2}{2}\frac{\Sigma R^2}{P^2 
Q^3},
\end{equation}

where the subscript $\lambda$ indicates that the partial derivative is being 
taken at fixed $\lambda$. \par
Inversion of the differential relations (\ref{drb}) and (\ref{dl_mu}) gives

\begin{equation}\label{dr_rb}
\Sigma R^2 dr=\Delta Q(d\rb -\mu P^2 d\lambda), \ \ \ \Sigma R^2d\theta 
= P (\Delta d\rb + \mu Q^2 d\lambda),
\end{equation}

which will be of use in subsequent sections. 

\section{Quasi-spherical light cones}

From here on we confine attention to a special class, ``quasi-spherical light 
cones'', defined as lightlike hypersurfaces which asymptotically approach 
Minkowski spherical light cones at infinity. Define an angle 
$\thetab(r,\theta)$ by 

\begin{equation}\label{l_def}
\lambda=\sin^2\thetab.
\end{equation}

In Minkowski space, the generators $\lambda=constant$ of light cones
are radial straight lines, suggesting (as we confirmed in Sec. 
\ref{case_m_0}) that $\thetab$ is the spherical polar angle. For $r>>a$, 
the oblate coordinates $(r,\theta)$ become indistinguishable from spherical 
coordinates, so we have the condition

\begin{equation}\label{thetas_ic}
\thetab(r=\infty, \theta)=\theta
\end{equation}

for spherical light cones in Minkowski space. This asymptotic condition 
must therefore also hold for quasi-spherical surfaces in Kerr space. This 
fixes the arbitrary function $f'(\lambda)$ in (\ref{F2}). We conclude that

\begin{equation}\label{Fqs}
F(r,\theta,\lambda,m)= \int_r^\infty \frac{dr'}{Q(r',\lambda,m)} - 
\int_\theta^\thetab \frac{d\theta'}{P(\theta',\lambda)}
\end{equation}

generates the solution for quasi-spherical surfaces. The radial function 
$\rho(r,\theta,\lambda,m)$ is given by (\ref{solution}) with (compare 
(\ref{g}) with $g(\lambda)=0$)

\begin{equation}\label{fl}
a f(\lambda)= -2 \int_0^{\thetab} \sqrt{\lambda-\sin^2\theta'} d\theta'.
\end{equation}

The equation of the hypersurface is then $v=t \pm \rb(r,\theta)=constant$, 
with $\rb=\rho(r,\theta,\lambda(r,\theta))$ and the function 
$\lambda(r,\theta) \equiv\sin^2\thetab$ determined by the constraint 
$F=0$. 

\section{No caustics for positive $r$}\label{no_caustic}

We now prove the rather remarkable result that quasi-spherical light cones 
are free of caustics for all positive values of the Kerr radial coordinate $r$. 
\par
This is trivially true if $m=0$, when these surfaces are simply light cones 
in Minkowski space with vertices at the spatial origin, represented by 
$r=0$, $\theta=0$ or $\pi$ in oblate spheroidal coordinates according to 
(\ref{oblate}). We shall prove that it is true \emph{a fortiori} for $m>0$ 
by effectively showing that when $m$ is larger, the generators converge 
\emph{less} rapidly as $r \rightarrow 0^+$. \par
As noted in Sec. \ref{gen_prop}, formation of a caustic along a generator 
is signalled by 

\begin{equation}\label{caust}
\mu P Q \sin\theta \rightarrow 0.
\end{equation}

We consider in turn the behaviour of each factor in (\ref{caust}) along an 
ingoing generator, so that $r$ is positive and decreasing, with $\lambda$ 
fixed. Because of the equatorial symmetry we need only consider 
``northern'' generators, i.e., we may assume $\thetab$ (which specifies the 
initial, asymptotic value of $\theta$ when $r=+\infty$) to be acute, and 
$P(\theta,\lambda)$ positive, at least initially. (Note from (\ref{drb}) that 
equatorial symmetry of $\rb(r,\theta)$ implies $P(\lambda, \pi-\theta)=-
P(\lambda,\theta)$.)\par
Since $\theta$ decreases with $r$ at fixed $\lambda$ according to 
(\ref{dl_mu}), the factor $P=a\sqrt{\lambda-\sin^2\theta}$ must 
\emph{increase} and remain positive. Any possible caustic in the Kerr 
positive-$r$ sheet cannot arise from the behaviour of $P$. \par
We proceed to consider the other factors in (\ref{caust}). From the 
definition (\ref{pq}),

\begin{equation}\label{qin}
Q(r,\lambda,m) > Q(r,\lambda,m=0) > 0\ \ \ (r > 0, m > 0).
\end{equation}

We shall further show that 

\begin{equation}\label{tiq}
(\partial\theta / \partial m)_{r,\lambda} >0 \ \ \ \mbox{and} \ \ \
(\partial \mu/ \partial m)_{r,\lambda} > 0 \ \ \
(r>0, \thetab < \pi/2).
\end{equation}

Thus, none of the three factors $\mu$, $Q$ and $\sin\theta$ can reach zero 
sooner for positive $m$ than they do in flat space, and hence they do not 
reach zero for any positive value of $r$. \par
To establish (\ref{tiq}), we note that the function $\theta(r,\lambda,m)$ is 
determined by the vanishing of $F(r,\theta,\lambda,m)$ as given by 
(\ref{Fqs}). Taking the differential at fixed $r$,$\lambda$ and using 
(\ref{pq}) gives

\begin{equation}\label{I}
(\partial\theta/\partial m)_{r,\lambda}=a^2\lambda P I, \ \ \
I(r,\lambda,m) \equiv \int_r^{\infty} r'dr'/Q^3(r',\lambda,m),
\end{equation}

which is manifestly positive. Turning to $\mu$, this is defined as a 
function of $r,\theta,\lambda,m$ by $\mu = -\partial F/ \partial \lambda$ 
according to (\ref{dl_mu}). Inverting the order of partial differentiation 
and using (\ref{Fqs}),

\begin{eqnarray}\label{du}
\frac{\partial\mu}{\partial m}= -\frac{\partial}{\partial \lambda}
\frac{\partial F}{\partial m}=\frac{\partial}{\partial \lambda}
(a^2 \lambda I), \\
\frac{\partial\mu}{\partial \theta} =- \frac{\partial}{\partial \lambda}
\frac{\partial F}{\partial \theta} = -\frac{\partial}{\partial \lambda}
\frac{1}{P}= \frac{a^2}{2 P^3}.
\end{eqnarray}

Hence 
\begin{equation}\label{du_dm}
\left(\frac{\partial\mu}{\partial m} \right)_{r,\lambda} = 
\frac{\partial\mu}{\partial m} + \frac{\partial\mu}{\partial \theta}
\left(\frac{\partial\theta}{\partial m} \right)_{r,\lambda} = 
a^2 \left(1+ \frac{1}{2}\frac{a^2\lambda}{P^2} \right) I +
\frac{3}{2}a^4 \lambda J,
\end{equation}

where

\begin{equation}\label{J}
J(r,\lambda,m)=\int_r^\infty r' \Delta(r') dr'/Q^5(r',\lambda,m).
\end{equation}

Each term in (\ref{du_dm}) is manifestly positive. This completes the 
proof. \par
The fact that adding mass to the Kerr field should \emph{reduce} the 
focusing of ingoing radial light rays is at first sight paradoxical, but can be 
explained by the circumstance that the material source is not at the origin, 
but located on the singular equatorial ring $x^2+y^2=a^2$, $z=0$ (see 
(\ref{oblate})). The source has a peculiar, ``demi-pole'' structure \cite{bi}, 
made possible by the double-sheeted structure of the Kerr manifold: a ring 
of positive mass in the sheet $r>0$ is bonded to a ring of negative mass in 
the sheet $r<0$. Light rays heading inwards toward the origin in the sheet 
$r>0$ are deflected outward by the ring and defocused. Repulsive effects 
have become dominant by the time they pass through the disk $r=0:$ the 
rays are then refocused and form a caustic in the sheet $r<0$. This 
description is more than just hand-waving: the Keres Newtonian analogue 
model of the Kerr field \cite{ki} has a source structure and equations of 
motion closely resembling Kerr (apart from frame dragging), and displays 
precisely this behaviour. \par
\begin{figure} 
\includegraphics[viewport =0 375 500 750, clip] {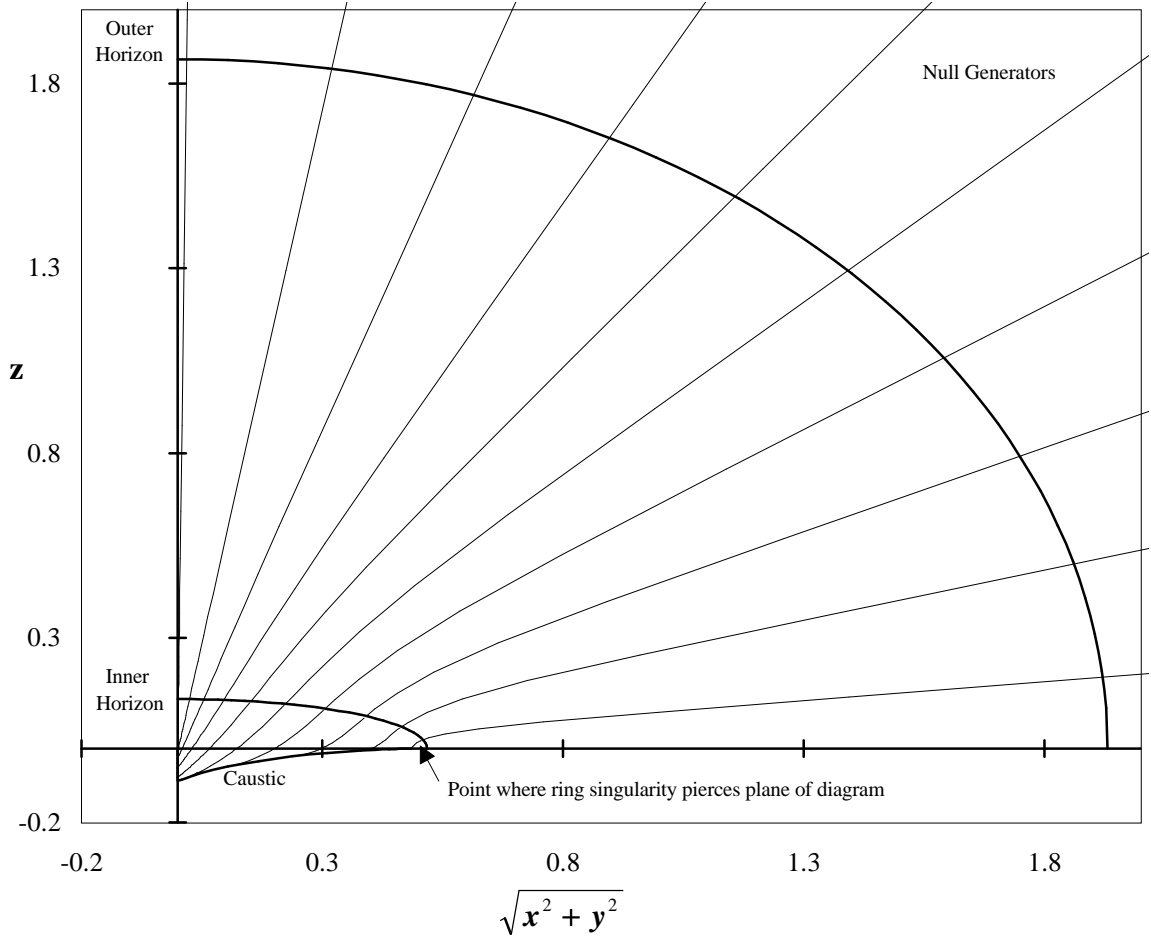}
\caption{\label{fig1}Null generators of the quasi-spherical light cones 
projected onto the Cartesian plane for $m=1$, $a=1/2$. The azimuthal 
twist (not shown in this two-dimensional representation) depends on the choice of azimuthal coordinate; it is zero for the coordinate $\phi_\epsilon$ introduced in Sec. 8. In order to follow the generators through the equatorial disk $r=0$ 
and into the negative-$r$ sheet, the ``southern half'' ($\theta > \pi/2$) of 
the positive-$r$ sheet has been cut away, and replaced by the ``northern'' 
half ($\theta < \pi/2$) of the negative-$r$ sheet.  Note that only well within 
the outer horizon does the mass of the spacetime significantly affect the 
$r,\theta$ behaviour of the generators (which would continue as straight 
lines intersecting the origin if $m$ were zero). The caustic emanates from 
the singularity $r=0$, $\theta=\pi/2$ (if $m>0$) and creeps inwards toward 
the axis and ``downwards'' toward increasingly negative values of $r$.} 
\end{figure}

Figure \ref{fig1} (a result of numerical integrations described in Sec. 10) 
shows the ingoing generators mapped onto the flat background of the 
Kerr-Schild decomposition, using the rectangular coordinates defined in 
(\ref{oblate}). (Effects of frame dragging are not shown). According to the 
Keres-Kerr model, the gravitational ``force'' should become repulsive for $ 
r \approx a$, and, indeed, the generators have inflection points near this 
radius. 

\section{Eikonal solution of the wave equation}
Since the quasi-spherical light cones $t\pm\rb=constant$ are 
characteristics of the wave operator, the coordinates $\rb,\lambda$ are well 
adapted for representing asymptotically spherical high-frequency 
modes.\par
The wave equation for $\Psi(\rb,\lambda,\phi,t)$ on the Kerr background 
(\ref{kerr_zamo}) is

\begin{eqnarray}\label{wave}
\frac{\Delta}{R^2} \BBox\Psi = (\partial^2_\rb -\partial^2_t)\Psi + 
(\partial_\rb \ln \gamma) (\partial_\rb \Psi) + \gamma^{-1} 
\partial_\lambda(\gamma^{-1} \Delta \sin^2\theta \partial_\lambda \Psi) + 
\\ \nonumber
+\left(1-\frac{2mr}{\Sigma}\right) \frac{1}{R^2\sin^2\theta} 
\partial^2_\phi \Psi - 2\omega_B \partial_\phi \partial_t \Psi =0
\end{eqnarray} 

where $ \gamma=\mu P Q \sin\theta$ gives the area element (degenerate 
volume element) on the light cone (\ref{dsll}).\par
Introducing the ansatz 

\begin{equation}\label{ansatz}
\Psi = \Phi(\rb,\lambda) e^{i\omega (t+\epsilon \rb)} e^{i n \phi} \ \ \ 
(\epsilon = \pm 1)
\end{equation}

into (\ref{wave}), we find

\begin{equation}\label{wave2}
2\epsilon i \omega \left[ \partial_\rb \Phi +\Phi \left(\frac{1}{2} \partial_\rb 
\ln\gamma - \epsilon i n \omega_B\right)\right] + B =0,
\end{equation}

where

\begin{equation}\label{B}
B=\partial^2_\rb \Phi + (\partial_\rb \ln \gamma)(\partial_\rb \Phi) + 
\gamma^{-1} \partial_\lambda(\gamma^{-1} \Delta \sin^2\theta 
\partial_\lambda \Phi) - \frac{n^2}{R^2\sin^2\theta}\left(1-
\frac{2mr}{\Sigma}\right) \Phi
\end{equation}

has no explicit $\omega$-dependence. We can re-express $\omega_B$ in 
(\ref{wave2}) as a partial derivative with respect to $\rb$. Defining an 
angular function 

\begin{equation}\label{alpha}
\alpha(r,\lambda)=\int^r \frac{2mar'}{\Delta(r')Q(r',\lambda)}dr'
\end{equation} 

and recalling (\ref{dr_rb}),we find 

\begin{equation}\label{wb2}
\left(\frac{\partial \alpha}{\partial \rb} \right)_\lambda = 
\frac{2mar}{\Delta Q} \left(\frac{\partial r}{\partial \rb} \right)_\lambda = 
\frac{2mar}{\Sigma R^2} = \omega_B.
\end{equation}

Therefore (\ref{wave2}) can be rewritten

\begin{equation}\label{wave3}
2\epsilon i \omega \partial_\rb (\sqrt\gamma e^{-\epsilon i n \alpha} \Phi) 
= -\sqrt\gamma e^{-\epsilon i n \alpha} B.
\end{equation}

If $\omega$ is large, we can apply an iterative procedure to (\ref{wave3}) 
and (\ref{B}) to develop the solution for $\Phi$ in inverse powers of 
$\omega$. In the lowest approximation, we simply equate the coefficient 
of $\omega$ in (\ref{wave3}) to zero. This yields the ``eikonal 
approximation'' 

\begin{equation}\label{eik}
\Psi \approx \gamma^{-1/2} f(\lambda) e^{i \omega (t+\epsilon \rb)} e^{i 
n (\phi +\epsilon \alpha)} 
\end{equation}

with an arbitrary function $f(\lambda)$. By linear superposition we can 
build from this an arbitrary high-frequency solution 

\begin{equation}\label{eik2}
\Psi \approx \gamma^{-1/2} F(\lambda, \phi +\epsilon \alpha, t+\epsilon 
\rb), 
\end{equation}

where $F$ is an arbitrary function of its three arguments varying rapidly 
with time. \par
From (\ref{eik}) and (\ref{wb2}) we see that crests of high-frequency 
azimuthal ($\phi$-dependent) waves twist about the axis with angular 
frequency $\omega_B$, the ZAMO angular velocity. Because 
$\omega_B$ and $\alpha$ depend on $\lambda$, propagation of azimuthal 
waves produces latitude-dependent phase shifts. The latitude-dependence 
is, however, always bounded, even when $\alpha \rightarrow \pm\infty$ at 
horizons. \par
In particular, wave-tails propagating inwards from the event to the Cauchy 
horizon of a spinning black hole experience a blueshift which is modulated 
by an oscillatory, latitude-dependent factor when $n\neq 0$ -- an effect 
first noted by Ori \cite{ori}. However, since $|n|$ is no larger than the 
multipole order $\ell$, this effect is more than offset by the natural power-
law decay of the higher multipole wave-modes with advanced time $v$,  
$\Psi \sim v^{-(2\ell+2)}$. It does not affect the uniformity of the leading-
order divergence $e^{|\kappa_-| v}$ of blueshift at the Cauchy horizon, 
where $\kappa_-$ (the inner-horizon surface gravity) is a constant. These 
features, which are expected to extend at least qualitatively to waves 
having (initially) lower frequencies, are important when considering the 
back-reaction of blueshifted wave-tails on the geometry near the Cauchy 
horizon \cite{pi}. 

\section {Kruskal coordinates} 

It is straightforward to transform the Kerr metric (\ref{kerr_zamo}) into a 
Kruskal-like form.
We introduce retarded and advanced times $u$ and $v$, and associated 
Kruskal coordinates $U$ and $V$, by the definitions 

\begin{equation}\label{dUV}
-\frac{dU}{\kappa U}=du=d(t-\rb), \ \ \ \frac{dV}{\kappa V}=dv=d(t+\rb).
\end{equation}

Here, $\kappa$ is the surface gravity of the horizon under consideration, 
defined for the outer ($r=r_+$) and inner ($r=r_-$) horizons by 

\begin{equation}\label{kappa}
\kappa_{\pm}=\pm\sqrt{m^2-a^2}/(r^2_\pm+a^2).
\end{equation}

Then (\ref{kerr_zamo}) becomes 

\begin{equation}\label{kerr_k}
ds^2=\frac{1}{\kappa^2R^2}\frac{\Delta}{UV}dUdV+\frac{L^2}{R^2}d
\lambda^2 + R^2\sin^2\theta (d\phi-\omega_B dt)^2
\end{equation}

with 

\begin{equation}\label{UV}
-UV=e^{2\kappa\rb} \ \ \ \mbox{and}  \ \ \ -V/U=e^{2\kappa t}.
\end{equation}

The first ($dUdV$) term is manifestly regular at the horizon sheets $U=0$ 
and $V=0$. The last term, involving the Boyer-Lindquist coordinates 
$\phi$ and $t$, is not. We therefore define the advanced and retarded 
angular coordinates $\phi_+$, $\phi_-$ by 

\begin{equation}\label{phi_ang}
\phi_\epsilon=\phi+\epsilon \alpha(r,\lambda) \ \ \ (\epsilon=\pm1)
\end{equation}

where $\alpha$ was defined in (\ref{alpha}). It is straightforward to show 
that

\begin{equation}\label{dalpha}
d\alpha=\omega_B d\rb -N d\lambda, \ \ \ N=\mu P^2 \omega_B + m a^3 I
\end{equation}

with $I$ defined as in (\ref{I}). Thus the last term in (\ref{kerr_k}) can be 
expressed in a manifestly regular way as

\begin{equation}\label{d_ang}
d\phi - \omega_B dt = d\phi_\epsilon - \omega_B du_\epsilon + \epsilon N 
d\lambda,
\end{equation}

with $u_+=v$, $u_-=u$. In (\ref{d_ang}) we can choose either sign for 
$\epsilon$, depending on which sheet of the horizon is of interest. 
For example, $\phi_+$ and $u_+$ are regular on the ``upward'' (future) sheets of both outer and inner horizons, with the nice bonus feature that $u_+$ is constant over each ingoing light cone and $\phi_+$ is constant along each ingoing generator. \par
Is there a single azimuthal coordinate which regularizes the metric simultaneously on both past and future sheets of (say) the outer horizon, including the bifurcation surface? Since $\omega_B$ takes a constant value $\omega_H=a/2mr_+$ over this horizon, a possible choice for the desired coordinate is $\Phi=\phi-\omega_H t$, and $\Phi$ in fact agrees with $\phi_+$ on the future sheet and $\phi_-$ on the past sheet. But $\Phi$ develops the undesirable features characteristic of rigidly rotating axes in the outer regions of the space.\par
We briefly mention that following standard methods one could construct a 
Penrose diagram of the Kerr spacetime using the Kruskal-like coordinates 
(\ref{UV}), valid up to formation of the caustic surface. Such a 
diagram would not look any different from the usual textbook examples 
\cite{di} except that a \emph{single} 2-dimensional diagram is all that 
would be needed to illustrate the causal structure of the spacetime. This is 
because the ingoing ($V=constant$) and outgoing ($U=constant$) lightlike 
congruences intersect at surfaces of constant $\rb$, represented by a single 
point on a diagram where the compactified coordinate system is derived 
from (\ref{UV}). The causal future of observers on the surface of 
intersection is entirely contained within the future-directed wedge of 
$U=constant$, $V=constant$ (as is evident from (\ref{kerr_zamo}) by 
noting that when $|d\rb| > |dt|$ the line-element is spacelike). 

\section{Limiting cases and approximations}
The quasi-spherical coordinates $\rb$ and $\lambda=\sin^2\thetab$, and 
the coefficients in the quasi-spherical forms (\ref{kerr_zamo}) and 
(\ref{kerr_k}) of the Kerr metric are complicated elliptic functions of the 
Kerr-Boyer-Lindquist coordinates $r$,$\theta$. It may therefore be useful 
to record here the simple forms these functions reduce to in the two 
opposite limiting cases, $m/a \rightarrow 0$ and $m/a \rightarrow \infty$. 
\par
(i) $m=0$, $a \neq 0$: This was studied in Sec. 3, and we simply list the 
results.

\begin{equation}\label{limf1}
\rb=\sqrt{r^2+a^2\sin^2\theta}, \ \ \ \tan\thetab =\frac{\sqrt{r^2+a^2}}{r} 
\tan\theta,
\end{equation}

\begin{equation}\label{limf2}
Q=\frac{r(r^2+a^2)}{\sqrt{r^2+a^2\sin^2\theta}}, \ \ \ P=\frac{a^2 
\sin\theta \cos\theta}{\sqrt{r^2+a^2\sin^2\theta}}, \ \ \ \mu 
P=\frac{1}{2}\frac{\sin\theta \cos\theta}{(\sin\thetab \cos\thetab)^2},
\end{equation}

\begin{equation}\label{limf3}
\gamma=\frac{1}{2} \frac{r^2+a^2\sin^2\theta}{\cos\thetab}.
\end{equation}

(ii) $m \neq 0$, $a=0$ : This is just Schwarzschild spacetime, and one 
readily finds 

\begin{equation}\label{lims1}
\rb=\int \frac{dr}{1-2m/r}, \ \ \ \thetab=\theta, \ \ \ Q=r^2, \ \ \ \mu 
P=\frac{1}{\sin 2\theta},
\end{equation} 

\begin{equation}\label{lims2}
\gamma=\frac{1}{2}\frac{r^2}{\cos\theta}.
\end{equation}

Comparison with numerical results described in the next section suggest that the $m=0$, $a \neq 
0$ case functions can provide serviceable analytic approximations outside of the outer horizon 
for broad ranges of $m$ and $a$. For example, with $m=1$ and $a=1/2$, all of the quantities 
except $\rb$ in (\ref{limf1}), (\ref{limf2}) and (\ref{limf3}) differ by no more than roughly 1 to 
3 percent from the true (numerically integrated) values, outside of $r_+$. ($\rb$ converges only 
logarithmically to the Minkowski value in the limit $r \rightarrow \infty$, and of course diverges 
at the horizon). 

\section{Numerical evolution of the generators}
In Figure \ref{fig1} we illustrated the behaviour of the generators and the 
nature of the caustic that forms in the negative $r$ sheet of the Kerr 
manifold. The $\lambda=constant$ curves were obtained by numerically 
integrating the evolution equations of the null generators of the 
hypersurface: 

\begin{equation}\label{evo}
\ell^{\alpha} = dx^{\alpha}/d\tau = - g^{\alpha \beta}\partial_{\beta} v,
\end{equation}

with affine parameter $\tau$. In particular

\begin{equation}\label{rtdot}
\dot{r}  \equiv  dr/d\tau=-\epsilon Q/\Sigma,  
\ \ \dot{\theta} \equiv d\theta/d\tau = -\epsilon P/\Sigma.
\end{equation}

To detect the formation of a caustic (see \ref{caust}) we track the 
evolution of $\mu$, and to reconstruct surfaces of constant $\rb$ in the 
$r,\theta$ plane we calculate the change of $\rb$ along the generators. If 
we treat $r$ as the independent variable, from 
(\ref{rtdot}), (\ref{solution}), (\ref{dl_mu}) and (\ref{d_mu}) we find

\begin{equation}\label{dt_dr}
(\partial \theta /\partial r)_{\lambda} = \frac{P}{Q},
\end{equation}

\begin{equation}\label{drb_dr}
(\partial \rb /\partial r)_{\lambda}=\Sigma R^2/ \Delta Q,
\end{equation}

and

\begin{equation}\nonumber
(\partial \mu /\partial r)_{\lambda}=\frac{a^2}{2}\frac{\Sigma R^2}{P^2 
Q^3}.
\end{equation} 

At the horizons $\rb$ and its derivative with respect to $r$ diverge. These 
divergences are coordinate singularities which can be avoided numerically 
by subtracting off the infinities that occur there. Thus define $\rt$ as

\begin{equation}\label{rt}
\rt=\rb - \frac{ln|r-r_+|}{2\kappa_+}
-\frac{ln|r-r_-|}{2\kappa_-},
\end{equation}

where $\kappa_+$[$\kappa_-$] is the surface gravity at the 
outer($r_+$)[inner($r_-$)] horizon
as defined in (\ref{kappa}). So we actually integrate $\rt$ along the 
generators and 
retrieve $\rb$ from the result using (\ref{rt}). \par
For asymptotic intitial conditions for $\mu$ and $\theta$, we use the 
relations listed in 
(\ref{limf1}) and (\ref{limf2}) for a sufficiently large initial $r$. Using 
(\ref{I}) and (\ref{du_dm}) one can show that
$\mu(r,\lambda,m)-\mu(r,\lambda,m=0) \approx 
(a^2\lambda m/8\sin\theta \cos\theta) (1/r^2)$ and $\theta(r,\lambda,m)-
\theta(r,\lambda,m=0) \approx (a^4\lambda m \sin\theta \cos\theta / 
4)(1/r^5)$ for large $r$ along a given generator. The initial condition for 
$\rb$ is arbitrary ($\rb+constant$ is still a solution to (\ref{pde})). \par

\begin{figure}
\includegraphics[viewport = 0 550 500 750,clip] {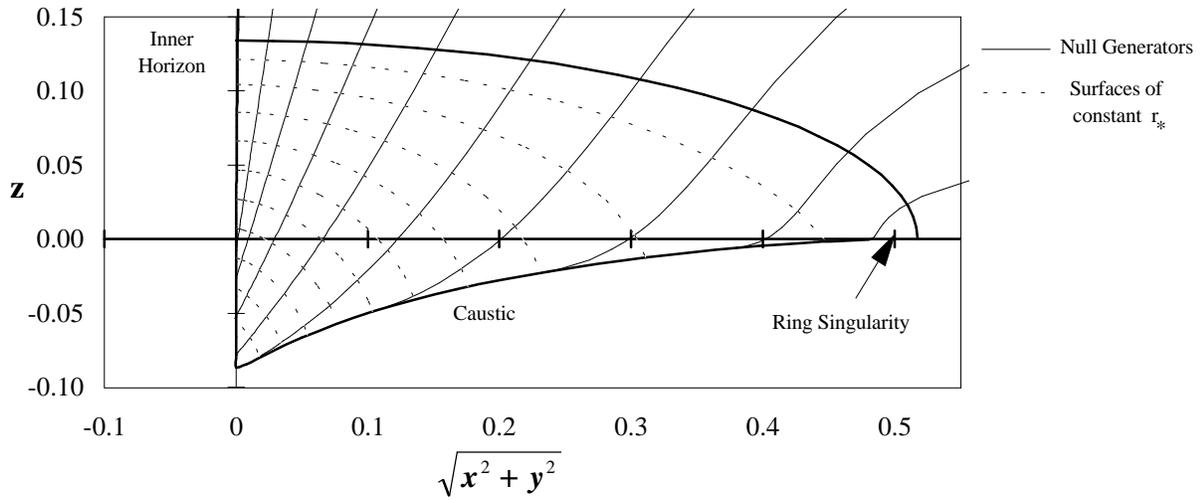}
\caption{\label{fig2}Null generators and surfaces of constant $\rb$ 
projected onto the Cartesian plane ($m=1, a=1/2$). This is a close-up of 
Figure \ref{fig1} focusing
on the region between the inner horizon and caustic. As in Figure 1 the 
northern half of the negative-$r$ sheet replaces the southern half of the 
positive-$r$ sheet to illustrate the evolution of the generators through the 
equatorial disk. 
}
\end{figure}

Figure \ref{fig2} below is a close-up within the inner horizon of the case 
illustrated in Figure \ref{fig1}, showing the projections of surfaces of 
constant $\rb$ onto the $\sqrt{x^2+y^2},z$ plane. This is the only region where $\rb = 
constant$ surfaces start showing significant deviations from the spheres of 
the massless scenario. Note that the caustic surface does not coincide with 
a hypersurface slice $\rb=constant$: following the generators inward a 
caustic first develops where the hypersurface meets the ring singularity, 
after which it quickly ``unravels''. We only present plots for $m=1$ and 
$a=1/2$; except for variations in scale there are no qualitative differences 
in the shapes of the curves and surfaces for arbitrary non-zero $a$ and 
positive $m$. 

\section{Concluding Remarks}
Despite the complexity of the coordinate transformations linking them to 
the familiar Boyer-Lindquist coordinates, the quasi-spherical coordinates 
$\rb$, $\thetab$ introduced in this paper and the light cones associated 
with them provide new and useful insights into the structure of the Kerr 
geometry and wave propagation in Kerr spacetime. We anticipate that they 
will find increasing use as their special advantages become apparent.

\vfill
\noindent
{\bf Acknowledgements} 
This work was supported by NSERC of Canada and by the Canadian 
Institute for Advanced Research.

\end{document}